\documentstyle[pra,aps]{revtex}

\begin{document}
\draft
\title{Comment on: rotational properties of trapped bosons}
\author{F.~Brosens and J.T. Devreese,}
\address{Departement Natuurkunde, Universiteit Antwerpen (UIA),\\
Universiteitsplein 1, B-2610 Antwerpen}
\author{L. F. Lemmens,}
\address{Departement Natuurkunde, Universiteit Antwerpen (RUCA),\\
Groenenborgerlaan 171, B-2020 Antwerpen}
\date{September 2, revised version November 4, 1996.}
\maketitle

\begin{abstract}
Based on the Hellman-Feynman theorem it is shown that the average square
radius of a cloud of interacting bosons in a parabolic well can be derived
from their free energy. As an application, the temperature dependence of the
moment of inertia of non-interacting bosons in a parabolic trap is
determined as a function of the number of bosons. Well below the critical
condensation temperature, the Bose-Einstein statistics are found to
substantially reduce the moment of inertia of this system, as compared to a
gas of ``distinguishable'' particles in a parabolic well.
\end{abstract}

\pacs{PACS: 03.75.Fi, 05.30.Jp,32.80.Pj}

Herewith we repost our paper cond-mat/9611090 (1996). It was published in
Phys. Rev. A {\bf 55}, 2453 (March 1997), three years before
cond-mat/0003471 (2000) by Schneider and Wallis.

\end{document}